\documentclass[%
 reprint,
showkeys, 
amsmath,amssymb,
aps,
]{revtex4-1}

\usepackage{graphicx}
\usepackage{dcolumn}
\usepackage{bm}
\usepackage{circuitikz} 
\usepackage{natbib}
\usepackage{mathtools}
\usepackage{siunitx}

\begin{document}

\preprint{APS/123-QED}

\title{Coarse-grained lattice protein folding on a quantum annealer}

\author{Tomas Babej}
 \email{All correspondence to tomas@proteinqure.com}
 \affiliation{ProteinQure Inc., Toronto, Canada}
\author{Mark Fingerhuth}%
 \affiliation{ProteinQure Inc., Toronto, Canada}
 \affiliation{University of KwaZulu-Natal, Durban, South Africa}
\author{Christopher Ing}%
 \affiliation{ProteinQure Inc., Toronto, Canada}
\date{\today}

\begin{abstract}
Lattice models have been used extensively over the past thirty years to examine the principles of protein folding and design.
These models can be used to determine the conformation of the lowest energy fold out of a large number of possible conformations.
However, due to the size of the conformational space, new algorithms are required for folding longer proteins sequences.
Preliminary work was performed by \citet{babbush2012} to fold a small peptide on a planar lattice using a quantum annealing device.
We extend this work by providing improved Ising-type Hamiltonian encodings for the problem of finding the lowest energy conformation of a lattice protein.
We demonstrate a decrease in quantum circuit complexity from quadratic to quasilinear in certain cases.
Additionally, we generalize to three spatial dimensions in order to obtain results with higher correlation to the actual atomistic 3D structure of the protein and outline our heuristic approach for splitting large problem instances into smaller subproblems that can be directly solved with the current D-Wave 2000Q architecture.
To the best of our knowledge, this work sets a new record for lattice protein folding on a quantum annealer by folding Chignolin (10 residues) on a planar lattice and Trp-Cage (8 residues) on a cubic lattice.

\end{abstract}

\maketitle


\section{Introduction}
\label{sec:introduction}

Protein folding is an essential biological process wherein an extended protein chain spontaneously self-assembles into a three dimensional structure.
Folding typically involves a large-scale conformational change that is challenging to predict from the sequence of amino acids alone.
For this reason, researchers employ simplified representations of proteins and folding.
These methods enable the study of general properties of protein folding to verify analytical models, and are computationally tractable.
Lattice models are a simplified folding model which represents protein chains as self-avoiding walks on a grid.
This is a discrete, coarse-grained representation of the protein where each grid intersection can be occupied by a single amino acid, and the angle between consecutive amino acids is determined by the grid type (typically planar or cubic).
Lattice models were first employed in the context of protein folding by \citet{dill1985theory, lau1989lattice}.
The hydrophobic-polar (HP) model distinguishes between two classes of amino acids, hydrophobic (H) and polar (P), and was designed to capture that the hydrophobic effect is the main driving force for folding.
An alternative lattice model was proposed by \citet{miyazawa1985estimation}, the MJ model, which distinguishes between all twenty naturally occurring amino acids.
This model assigns different interactions strengths to each possible amino acid pair using a statistical potential obtained by the quasi-chemical approximation.

Improvements in computing power has made it possible for researchers to leverage the power of supercomputers such as Cray Titan \citep{craytitan} or Sunway TaihuLight \citep{sunwaytaihu} to build high-accuracy atomistic models of proteins.
All-atom computational methods, including physics-based and knowledge-based approaches, have provided useful insights into protein folding and design, but these models are still too computationally costly for high-throughput folding studies.
Molecular dynamics (MD) simulations are very resource intensive, and have only been utilized to fold small fast-folding proteins.
Knowledge-based methods, such as those utilizing fragment-based assembly or heuristic algorithms, cannot be used to recover a folding pathway and may require the ranking of thousands of candidate protein folds to determine the native state.
Lattice folding offers complimentary data to high-resolution models, which may assist in accelerating protein structure determination.
Specifically, a multiscale algorithm capturing both the high-level features of the folding energy landscape, such as preferred amino acid contacts near the native state, and the low-level capability of state-of-the-art physics and knowledge-based methods for refinement, may offer significant performance increases over high-resolution methods alone at a fraction of the computational cost.

Despite the advances seen in supercomputing hardware, such performance increases are not sufficient to overcome the scaling complexity of lattice protein folding.
Finding the minimum energy conformation of a lattice protein is a hard problem.
Even for the case of the two class HP model, researchers have shown that the problem of finding a minimum energy conformation of a lattice protein is NP-complete \cite{hart1997robust,unger1993finding,berger1998protein}.
Unless \texttt{P=NP}, this implies that there exists no classical algorithm that can find the lowest energy state of a lattice protein in polynomial time, and may restrict the use of lattice folding to small protein lengths (less than 100 amino acids).

In recent years, quantum annealers have been shown to be able to solve certain NP-hard problems more efficiently than classical computers \citep{mandra2017deceptive, king2017quantum, denchev2016computational}.
For this reason, quantum annealing performed on real-world experimental devices might provide solutions to the problem of finding the lowest energy conformation for lattice proteins of sizes unreachable for classical hardware.
In order to solve a problem on a quantum annealing device, the problem first needs to be encoded as an Ising-type Hamiltonian, whose ground state corresponds to the solution of the given problem.
Since the encoding determines the number of required variables as well as the connectivity between them, it is important to investigate different ways of encoding the problem.
\citet{perdomo2008construction} were first to develop a Hamiltonian encoding based on a binary grid coordinate system.
Subsequently, \citet{babbush2012construction} developed three novel ways of mapping lattice heteropolymer models into Ising-type Hamiltonians.
Finally, \citet{perdomo2012finding} picked the most resource efficient mapping (for the given protein length) and established a proof-of-concept by folding the protein PSVKMA on the D-Wave One processor.

In summary, to facilitate the construction of a hybrid algorithm for protein folding, we require the use of a quantum processor to obtain the lowest energy conformation of the lattice protein, circumventing the need to explore this vast conformational space classically.
Amino acid contacts from lattice folds are then provided as additional input for high accuracy atomistic modeling using classical computers.
For this, we employ restraint-based MD simulations in order enable faster simulation.
In this work, we focus on the problem of mapping the lattice protein models onto existing quantum annealing hardware and the application of these methods for folding proteins on the D-Wave 2000Q quantum annealer.


\section{Mapping proteins onto discrete lattices}
\label{sec:mapping_lattice_proteins}

In this work, we build upon the methodology described by \citet{babbush2012construction} to encode the lattice folding problem into an Ising-type Hamiltonian.
Due to their low correlation with actual protein structure, we move away from planar lattices and instead derive encodings for cubic lattices for three different mappings of the lattice protein folding problem.

Specifically, for the coarse-grained lattice models discussed in this paper we define the \textit{lattice protein fold} as a path in the graph of a given grid, which does not visit any vertex twice and assigns an amino acid type to each visited vertex.
The energy of the protein fold can be calculated as the sum of interaction energies between adjacent non-covalently bound amino acids defined by a contact potential.
Multiple contact potentials are described in the literature.
Two of the most commonly used potentials are the HP and the MJ interaction potentials.
The HP potential only differentiates between two types of amino acids, hydrophobic (H) and polar (P), and it assigns a negative (favorable) weight to interactions between two adjacent, non-covalently bound hydrophobic residues.
The MJ potential differentiates between all 20 naturally occurring amino acids, assigning varying negative (favorable) weights to interactions between adjacent amino acids.

In order to solve the lattice protein folding problem on a quantum annealing device which leverages Ising-type Hamiltonians, such as the D-Wave 2000Q, we first need to construct an injective mapping between the set of all possible lattice protein folds and the set of binary strings, represented by a sequence of qubits in the machine.
This allows us to uniquely decode a given solution string into a lattice protein fold.
Two different approaches are outlined in sections \ref{sec:binary_turn_encoding} and \ref{sec:binary_flag_encoding}.

The next step is to construct the energy landscape of for the Ising system such that the valid, lowest energy conformation of the lattice protein corresponds to the ground state of the system.
On the hardware level, this is realized by introducing couplings between Ising variables.
On the logical level, we use pseudo-boolean expressions that are subsequently reduced to 2-local interactions implementable on the device.
They encode the interaction potential (HP or MJ) and they prevent non-physical solutions such as overlapping or disconnected amino acid sequences.
There can be different ways of generating these pseudo-boolean expressions even using the same lattice fold encoding.
Each has its own advantages and disadvantages which will be discussed in their respective sections, see \ref{sec:turn_ancilla}, \ref{sec:turn_circuit} and \ref{sec:nested_shell_encoding}.

\subsection{Binary encoding of lattice folding into turns}
\label{sec:binary_turn_encoding}

\begin{figure}[h!]
	\centering \includegraphics[width=0.5\textwidth]{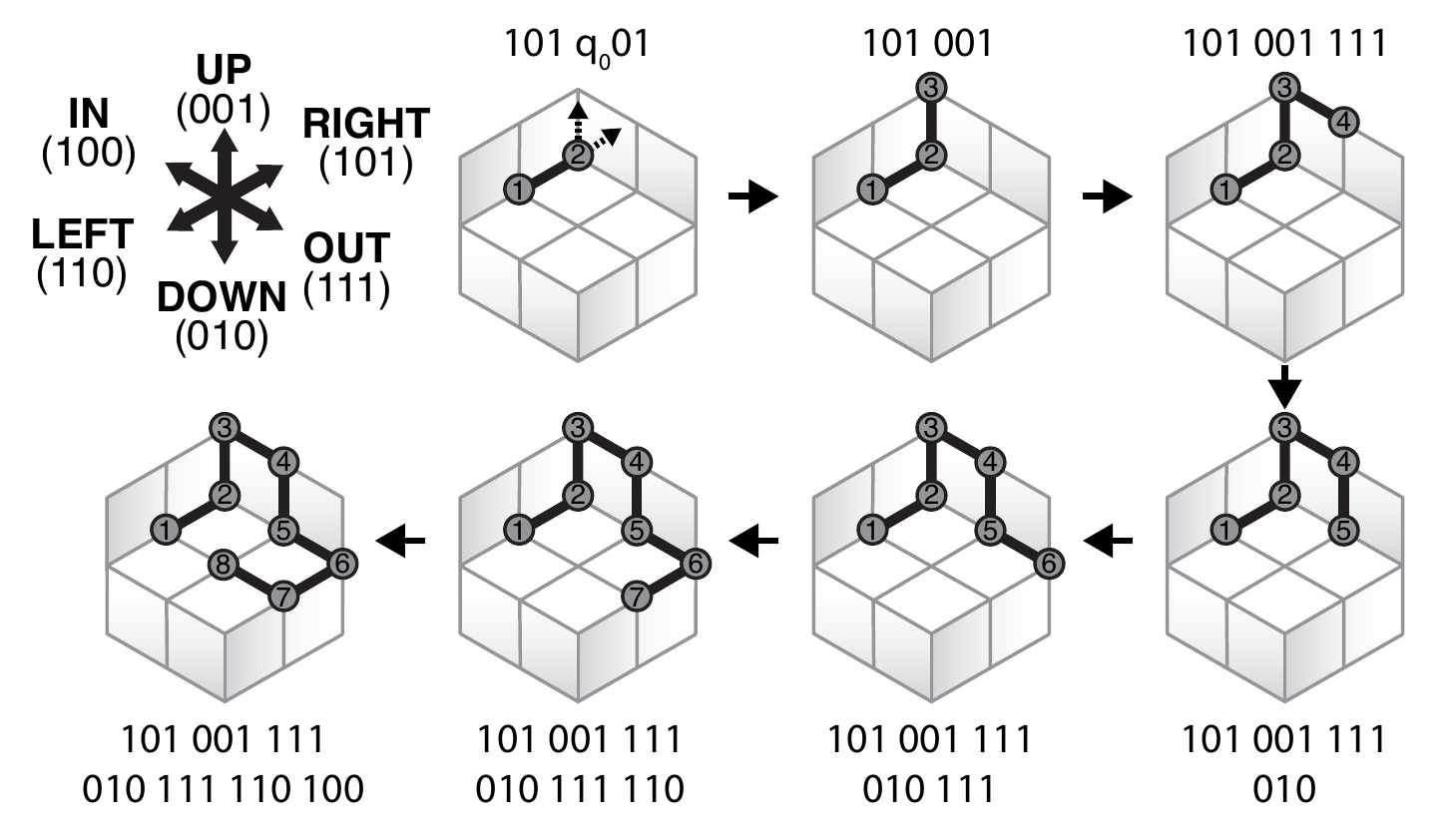}
	\caption{\label{fig:binary_turns_3d}Instead of using absolute coordinates, we define lattice proteins in binary through globally defined turns. On a cubic lattice every possible turn can be encoded into 3 binary bits as shown in the top left corner.}
\end{figure}

The most straightforward way of mapping lattice proteins to binary is to impose a binary coordinate system onto the cubic 3D lattice as it was first discussed by \citet{perdomo2008construction}.
However, in three dimensions this encoding is fairly inefficient, requiring $\Omega(N\log N)$ qubits to encode a lattice protein of length $N$.
Given the small size of today's quantum processing units one wants to be as efficient as possible when it comes to resources and for this reason we will not further discuss this encoding strategy in this whitepaper.
A more compact way of encoding lattice proteins on a cubic lattice is using globally defined directions called \textit{turns}.
Since a lattice protein is a path in the lattice graph, it is also specified as a sequence of edge directions, given that we fix the initial point as the point of origin.
Fig. \ref{fig:binary_turns_3d} shows a binary mapping that encodes each of the six spatial directions on a cubic lattice as 3 qubits, which requires a total of $\Omega(N)$ qubits.
Note that the bit strings for the directions have been carefully chosen in order to optimize the localility of pseudo-boolean expressions.

To decrease the number of required qubits even further, we fix the first three qubits to enforce a 'right' move and due to rotational symmetry we can fix two additional qubits in the second turn.
The solution bit string is then given by:

\begin{equation}
	\label{equ:turn_bitstring}
	\mathbf{q} = \underbrace{101}_\text{turn 0}\: \underbrace{q_001}_\text{turn 1}\: \underbrace{q_1q_2q_3}_\text{turn 2} ... \underbrace{q_{(3N-11)}q_{(3N-10)}q_{(3N-9)}}_\text{turn $N-2$}.
\end{equation}

This binary encoding scheme is used for the turn ancilla as well as the turn circuit encoding in the following two sections.

\subsubsection{Turn ancilla encoding}
\label{sec:turn_ancilla}

This encoding introduces ancillary qubits into the Hamiltonian to encode information about amino acid interactions.
This ensures that the k-locality of the resulting Hamiltonian is bounded, unlike in the case of the turn circuit encoding (see Section \ref{sec:turn_circuit}).
Restricting the k-locality is critical, since all experimental devices, including D-Wave's quantum annealer only allow for 2-local interactions, and conversion of the Hamiltonian from k-local to 2-local comes at the cost of introducing additional ancilla qubits.
More specifically, the turn ancilla mapping requires:

\begin{align}
	(3N - 8) &+ \sum^{N-5}_{i=0} \sum^{N-1}_{j=i+4} \lceil 2\log_2(i-j)\rceil \big\{ (1+i-j)\,\text{mod}\,2 \big\} \nonumber \\
&+ \sum^{N-4}_{j=0} \sum^{N-1}_{k=j+3} \big[ (j-k)\,\text{mod}\,2 \big]
\end{align}

qubits for a protein of length N.

The Hamiltonian constructed with the turn ancilla scheme consists of the following four subcomponents,
\begin{equation}
H(\mathbf{q}) = H_\text{back}(\mathbf{q}) + H_\text{redun}(\mathbf{q}) + H_\text{olap}(\mathbf{q}) + H_\text{pair}(\mathbf{q}).
\end{equation}

The first component, $H_\text{back}$ penalizes lattice protein folds in which two consecutive edges go between the same pair of vertices, that is, edge $(v_1, v_2)$ followed by the edge $(v_2, v_1)$.
This fold is not valid since the protein goes back on itself which is, therefore, penalized to ensure that the ground state solution does not have this property.
The second component $H_\text{redun}$ penalizes the occurence of the two redundant 3-bit strings (which do not encode any valid direction on the grid).
The third component $H_\text{olap}$ penalizes folds in which any residues $i$ and $j$ with $j > i+3$ occupy the same grid point, complementing $H_\text{back}$.
All penalties in the first three terms are adding large positive numbers to the overall energy such that these solutions do not lie close to the ground state.
The components $H_\text{back}$ and $H_\text{olap}$ could technically be expressed as one component, but treating them separately reduces the number of required ancillary qubits.
The last component $H_\text{pair}$ accounts for the interaction between non-bonded amino acids that are adjacent on the lattice using the HP or MJ interaction potential.
Moving away from planar lattices \cite{babbush2012construction}, all of our derivations are performed for the case of cubic lattices which are much better suited given their higher correlation to the atomistic tertiary structure.

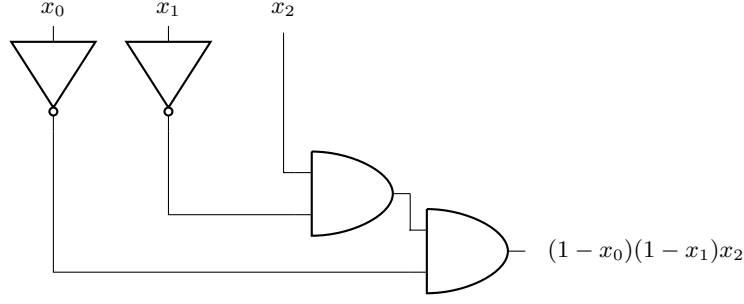
\begin{figure*}
\begin{circuitikz}[scale=0.765]
\draw

(0,5.2) node[] (x0) {$x_0$}
(2,5.2) node[] (x1) {$x_1$}
(4,5.2) node[] (x2) {$x_2$}
(10.3,1) node[] (output) {$(1-x_0)(1-x_1)x_2$}

(0,4) node[american not port, draw ,rotate=-90] (not0) {}
(2,4) node[american not port, draw ,rotate=-90] (not1) {}
(6,2) node[and port, draw] (and0) {}
(8,1) node[and port, draw] (and1) {}

(not1.out) |- (and0.in 2)
(4,4.8) |- (and0.in 1)
(not0.out) |- (and1.in 2)
(and0.out) |- (and1.in 1);

\end{circuitikz}
\caption{\label{fig:boolean_001_true}Classical boolean circuit representing an 'up' move on the cubic lattice. The circuit evaluates to TRUE (equals to 1) if and only if $x_0x_1x_2=001$}
\end{figure*}

\textbf{Construction of $H_\text{back}$.} For the construction of the pseudo-boolean expression we need to derive helper boolean logic circuits.
More specifically, to correctly parse every possible turn in the bit string, we need to construct boolean circuits that yield TRUE given exactly one of the valid 3-bit strings that encode directions, and FALSE otherwise.
For example, the bit string $001$ corresponds to an 'up' ($+y$) move on the cubic lattice and the boolean circuit that yields TRUE for this input is shown in Fig.~\ref{fig:boolean_001_true}.
Note that any other input to this circuit will output FALSE.

Using these boolean circuits as building blocks, we arrive at six closed-form expressions for the six spatial directions 'right' ($+x$), 'left' ($-x$), 'up' ($+y$), 'down' ($-y$), 'in' ($-z$) and 'out' ($+z$).
From there, it is straightforward to derive six functions that evaluate to TRUE if and only if the $j$-th turn goes into the respective direction:
\begin{align}
	\label{eq:right_true}
	d^j_{+x} &= (1-q_{2+\phi})q_{1+\phi}q_{3+\phi}, \\
	\label{eq:left_true}
	d^j_{-x} &= (1-q_{3+\phi})q_{1+\phi}q_{2+\phi},  \\
	\label{eq:up_true}
	d^j_{+y} &= (1-q_{1+\phi})(1-q_{2+\phi})q_{3+\phi}, \\
	\label{eq:down_true}
	d^j_{-y} &= (1-q_{1+\phi})(1-q_{3+\phi})q_{2+\phi}, \\
	\label{eq:out_true}
	d^j_{+z} &= q_{1+\phi}q_{2+\phi}q_{3+\phi}, \\
	\label{eq:in_true}
	d^j_{-z} &= (1-q_{2+\phi})(1-q_{3+\phi})q_{1+\phi},
\end{align}

where $\phi=3(j-2)$ for the sake of simplicity.
Even though there are only six spatial directions on a cubic lattice, there is eight unique 3-bit strings.
Since we encode the turns of the protein into 3-bit strings these two additional 3-bit strings will unavoidably be part of the solution space as well, even though they do not correspond to any move.
In order to penalize the occurence of the these strings later on, we derive the two functions that check if the $j$-th move is described by one of the two invalid bit strings $000$ or $011$:
\begin{align}
	\label{equ:000_invalid}
	d^j_{000} &= (1-q_{1+\phi})(1-q_{2+\phi})(1-q_{3+\phi}),&& \\
	\label{equ:011_invalid}
	d^j_{011} &= (1-q_{1+\phi})q_{2+\phi}q_{3+\phi}.&&
\end{align}

The resulting subcomponent is then given as:
\begin{align}
	H_\text{back}(\mathbf{q}) \:&= \:\lambda_\text{back} \Big\{ (q_0 \land d^{\,2}_{-x}) + ((1-q_0) \land d^{\,2}_{-y}) \nonumber\\
						   &+ \sum^{N-3}_{j=2} \big[ (d^j_{+x}\land d^{j+1}_{-x}) + (d^j_{-x}\land d^{j+1}_{+x}) \nonumber\\
						   &\quad\quad + (d^j_{+y}\land d^{j+1}_{-y}) + (d^j_{-y}\land d^{j+1}_{+y}) \nonumber \\
						   &\quad\quad + (d^j_{+z}\land d^{j+1}_{-z}) + (d^j_{-z}\land d^{j+1}_{+z}) \big] \Big\}.
\end{align}

\textbf{Construction of $H_\text{redun}$.} This part of the Hamiltonian imposes an energy penalty $\lambda_\text{redun}$ in case one of the two invalid 3-bit strings, $000$ or $011$, occurs in a turn since these two strings do not encode any direction on the cubic lattice.
Eq.~\ref{equ:000_invalid} and \ref{equ:011_invalid} are functions that equal to TRUE when the $j$-th turn is an invalid move.
The way the first two turns are constructed in Eq.~\ref{equ:turn_bitstring} (containing explictly fixed qubits) do not allow for an invalid move so $H_\text{redun}$ does only apply starting from turn 2.
Thus, $H_\text{redun}$ can be written as,
\begin{equation}
	H_\text{redun} = \lambda_\text{redun} \sum^{N-2}_{j=2} \big( d^j_{000} + d^j_{011} \big) .
\end{equation}

\textbf{Construction of $H_\text{olap}$.} In order to to penalize more complex overlaps e.g.
the $i$-th amino acid overlapping with the $(i$$+$$6)$-th amino acid, one needs to be able to derive the coordinates of the respective amino acids relative to the first amino acid in the chain from the information about the turns.
Since Eq.~\ref{eq:right_true}-\ref{eq:in_true} keep track of the direction at each turn, the following three position functions can be constructed:
\begin{align}
	x_m &= \begin{cases}
		0, & \text{if m=0},\\
		1 + q_0 + \sum^{m-1}_{j=2} \big( d^j_{+x} - d^j_{-x}\big) , & \text{otherwise}, \\
	\end{cases} \\\
	y_m &= \begin{cases}
		0, & \text{if m=0},\\
		1-q_0 + \sum^{m-1}_{j=2} \big( d^j_{+y} - d^j_{-y}\big), & \text{otherwise}, \\
	\end{cases} \\
	 	z_m &=  \sum^{m-1}_{j=2} \big( d^j_{+z} - d^j_{-z}\big).
\end{align}

Note, that the we needed two cases since the zero-th amino acid constitutes a special case - it occupies the origin.
Furthermore, the first terms in $x_m$ and $y_m$ with $m>0$ are due to the fixed qubits defined in $\mathbf{q}$ from Eq.~\ref{equ:turn_bitstring}.
In order to determine the distance between any two amino acids we define the squared distance on a cubic lattice as,
\begin{equation}
	D_{jk} = (x_j - x_k)^2 + (y_j - y_k)^2 + (z_j - z_k)^2 .
\end{equation}

$D_{jk}$ is zero if the $j$-th and $k$-th amino acid occupy the same point on the lattice and positive otherwise.
Furthermore, due to the way we mapped the 3-bit strings onto the six directions $D_{jk}$ has the beneficial property of being at most 4-local.
As given in \cite{babbush2012construction}, the bounds of $D_{jk}$ are
\begin{equation}
	0 \leq D_{jk} \leq (j-k)^2 .
\end{equation}

In $H_\text{olap}$ we want to penalize non-trivial overlaps and, therefore, ensure that $D_{jk} \neq 0, j>k+3$.
This means, we want to enforce the inequality constraint $D_{jk} \geq 1$.
In order to transform an inequality into an equality we introduce the slack variable $\alpha_{jk}$:
\begin{equation}
	\label{equ:slack_bounds}
	0 \leq \alpha_{jk} \leq (j-k)^2 -1 .
\end{equation}
From this,
\begin{equation}
	\label{equ:slack_inequality1}
	\forall\, D_{jk} \geq 1 \:\exists\, \alpha_{jk}: (j-k)^2 - D_{jk} - \alpha_{jk} = 0 ,
\end{equation}
follows immediately.
And we observe that in the case $D_{jk}=0$ the following holds true,
\begin{equation}
	\label{equ:slack_inequality2}
	(j-k)^2 - D_{jk} - \alpha_{jk} \geq 1 \forall \, \alpha_{jk} .
\end{equation}


The important step in the turn ancilla encoding is how this slack variable is entering the Hamiltonian.
First proposed in \cite{babbush2012construction}, we will encode the value of $\alpha_{jk}$ in binary using ancillary qubits.
Ancilla qubits are additional unconstrained qubits that are used for intermediate computations but ultimately disregarded in the final solution string.

The number of amino acid pairs that require a slack variable is given by,
\begin{equation}
	N_{pairs} = \sum^{N-5}_{i=0} \sum^{N-1}_{j=i+4} \big[ (1+i-j)\,\text{mod}\,2\big].
\end{equation}

Each slack variable $\alpha_{jk}$ requires $\mu_{jk}$ ancillas to be stored.
The expression for $\mu_{jk}$ can be written as,
\begin{equation}
	\mu_{jk} = \lceil 2\log_2(j-k)\rceil \big\{ (1+j-k)\,\text{mod}\,2 \big\}.
\end{equation}

Thus, we can compute the total number of required ancillary qubits as,
\begin{equation}
	N_{ancilla} = \sum^{N-5}_{i=0} \sum^{N-1}_{j=i+4} \mu_{ij}
\end{equation}

Finally, using big-endian format we can obtain the value of the slack variable $\alpha_{jk}$ as,
\begin{equation}
	\alpha_{jk} = \sum^{\mu_{jk}-1}_{k=0} q_{p_{jk}+k}2^{\mu_{jk}-1-k} \, ,
\end{equation}
where $p_{jk}$ is a pointer that yields the index of the first ancilla encoding $\alpha_{jk}$.
In our case, we chose to attach the ancilla register to the end of the solution string $\mathbf{q}$ (Eq.~\ref{equ:turn_bitstring}) and thus the pointer is given by,

\begin{align}
	p_{jk} = (3N-8) + \sum^{j}_{u=0} \sum^{N-1}_{n=u+4} \mu_{un} - \sum^{N-1}_{m=k} \mu_{jm}.
\end{align}
As shown in \cite{babbush2012construction}, the value of the slack variable does not fall within the range defined in Eq.~\ref{equ:slack_bounds} but rather into the range,
\begin{equation}
0 \leq \alpha_{jk} \leq 2^{\mu_{jk}} - 1 .
\end{equation}

Therefore, we need to adjust Eq.~\ref{equ:slack_inequality1} and \ref{equ:slack_inequality2} accordingly,
\begin{align}
	\label{equ:adjusted_inequality1}
	\forall\: D_{jk} \geq \:1 \:\exists \:\alpha_{jk}: \:2^{\mu_{jk}} -  D_{jk} - \alpha_{jk} = 0 , \\
	2^{\mu_{jk}} -  D_{jk} - \alpha_{jk} \geq 1\: \forall \:\alpha_{jk} .
\end{align}

We want to ensure that $\alpha_{jk}$ takes the correct value such that Eq.~\ref{equ:adjusted_inequality1} equals to 0 if $D_{jk} \geq 1$.
Additionally, we want to avoid $\alpha_{jk}$ from taking negative values since this would lead to a decrease in energy and, thus, favour overlaps.
For this reason, we square the expression to ensure the correct behaviour and multiply it with a penalty $\lambda_\text{olap}$:
\begin{equation}
	\gamma_{jk} = \lambda_\text{olap} \big[ 2^{\mu_{jk}} - D_{jk} - \alpha_{jk} \big]^2 .
\end{equation}

Now everything is in place to construct the expression,
\begin{equation}
	H_\text{olap}(\mathbf{q}) = \sum^{N-5}_{i=0} \sum^{N-1}_{j=i+4} \big[(1+i-j)\,\text{mod}\,2 \big] \gamma_{ij}
\end{equation}
where the term $\big[(1+i-j)\,\text{mod}\,2 \big]$ eliminates unneccessary terms since only amino acids that are an even number of turns away from each other can overlap.

\textbf{Construction of $H_\text{pair}$.} This final component of the turn ancilla encoded Hamiltonian models the interaction between non-covalently bonded neighbouring amino acids on the lattice.
In order to do so, we need to construct an interaction matrix $P$ using either the HP model by \citet{lau1989lattice} or the MJ potential by \citet{miyazawa1985estimation}.
The HP model is rather simple - the only interaction coefficient is $-1$ in the case of two adjacent hydrophobic amino acids.
For this reason, it yields slightly smaller Hamiltonians which - in some cases - can enable solving larger problem instances.
The MJ model assigns different interaction strengths to each pairwise interaction of all 20 amino acids and hence models protein folding much more accurately.
Note that $P$ differs from protein to protein and is not equal to the MJ matrix even though it is constructed using its elements.
Given the fact that a lot of amino acids will not be able to interact with each other due to parity reasons, $P$ is rather sparse.
For each possible interaction between residues we need to add one ancillary qubit flag $\omega_{jk}$ that is equal to 1 if and only if the $j$-th and $k$-th amino acid interact and 0 otherwise:
\begin{equation}
	\omega_{jk} =
	\begin{cases}
		1, & \text{if $D_{jk}=1$}.\\
    0, & \text{otherwise}.
  \end{cases}
\end{equation}


From this follows the interaction term,
\begin{equation}
	\vartheta_{jk} = \omega_{jk}P_{jk}(2-D_{jk}) .
\end{equation}

Summing over all possible interactions yields the final expression,
\begin{equation}
	H_\text{pair}(\mathbf{q}) = \sum^{N-4}_{j=0} \sum^{N-1}_{k=j+3} \big[ (j-k)\,\text{mod}\,2 \big] \omega_{jk}P_{jk}(2-D_{jk}),
\end{equation}

where we refined the approach by including the term $\big[ (j-k)\,\text{mod}\,2 \big]$.
This keeps the number of required ancillas as low as possible since amino acids that are separated by an even number of turns can not interact.

\subsubsection{Turn circuit encoding}
\label{sec:turn_circuit}

This encoding is the most qubit efficient mapping since it requires only $3N-8 \in O(N)$ qubits (and no additional ancilla qubits) to encode a protein of length N.
However, it comes at the disadvantage of containing many-body terms (involving more than two variables).
This poses a difficulty for the problem to be embedded on an experimental device, since those usually implement only 2-body interactions.

The k-body Hamiltonian can be reduced to a 2-body Hamiltonian with equivalent ground state by introducing ancilla qubits as resource efficient gadgets \cite{babbush2013resource}, hence, reducing the k-locality of the terms at the cost of increasing the size of the problem graph.
Yet, on small problem instances this encoding is more resource efficient than the other encodings described in this paper.
Below we give an asymptotically more efficient implementation of the turn circuit encoding of the cubic lattice protein folding problem than previously published in the literature \citep{babbush2012construction}.

The turn circuit encoded Hamiltonian consists of two main terms,
\begin{equation}
	H(\mathbf{q}) = H_\text{olap}(\mathbf{q}) + H_\text{pair}(\mathbf{q}) .
\end{equation}

\textbf{Sum strings.} Both subcomponents in the turn circuit encoded Hamiltonian, require us to introduce the notion of a 'sum string'.
In this case, the $i,j$-th $+x$ direction sum string represents the sum of the directional $+x$-strings (Eq.~\ref{eq:right_true}) from the $i$-th to the $j$-th amino acid.
By constructing sum strings for every pair of amino acids, we can keep track of the position of every amino acid on the lattice, which will enable us to test for possible overlaps or interactions of residues.

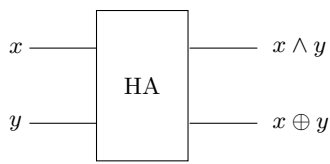
\begin{figure}[!b]
\begin{tikzpicture}

  \node[draw, minimum width=0.5cm, minimum height=2cm, text width=1cm, align=center] (HA) at (1,4) {HA};

  \draw[-] (-0.5,4.5) -- (0.39,4.5);
  \draw[-] (-0.5,3.5) -- (0.39,3.5);
  \draw[-] (1.63,3.5) -- (2.53,3.5);
  \draw[-] (1.63,4.5) -- (2.53,4.5);

  \draw (-1,4.5) node[label=right:$x$]{};
  \draw (-1,3.5) node[label=right:$y$]{};
  \draw (2.5,4.5) node[label=right:$x\land y$]{};
  \draw (2.5,3.5) node[label=right:$x \oplus y$]{};
\end{tikzpicture}
\caption{\label{fig:single_HA} A simple half adder circuit that takes two binary numbers $x$ and $y$ as input and results in their addition (mod 2) and the carry bit.}
\end{figure}

The sum strings are computed using half-adder circuits.
A half-adder circuit is shown in Fig.~\ref{fig:single_HA}.
It takes two binary numbers $x$ and $y$ as input and outputs two bits, the first one representing the carry bit $c = x \land y$ and the second one being the sum $s= x \oplus y$ which is equivalent to computing $(x+y)\,\text{mod}\,2$.
Hence, the half-adder results in the two-bit sum of $x$ and $y$.
As outlined in \cite{babbush2012construction}, larger half-adder circuits can be constructed to sum multiple input bits.

\begin{figure}[!t]
	\includegraphics[width=0.5\textwidth]{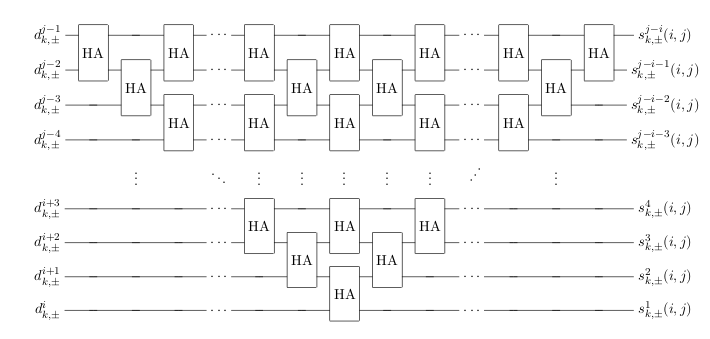}
    \caption{\label{fig:ha_circuit_babbush}Inefficient concatenation of half-adders to sum a sequence of bits \cite{babbush2012construction}.}
\end{figure}

Half-adder circuit can used to add up the directional bit strings between amino acid $i$ and $j$ in the $\pm k$ direction as shown in Fig.~\ref{fig:ha_circuit_babbush}.
However, this approach of building the circuit is inefficient in the number of half-adders required.
Below we propose a better circuit design that reduces the number of terms in the Hamiltonian.

The critical realization is that a binary sum string for two amino acids $j$ and $k$ requires at most of $\lceil\log_2(j-k)\rceil$ bits to encode the resulting value, where $1 \leq j < k \leq N$ and $j+1<k$.
This is important to note because the circuit shown in Fig~\ref{fig:ha_circuit_babbush} contains $(j-k)$ input as well as output bits.
Since a binary representation of $n$ uses at most $\lceil\log_2 n\rceil$ bits, it follows that $n - \lceil\log_2 n \rceil$ bits are not required to represent the sum string.
Using this insight we remove half-adders from the upper right section of the circuit once their information content is fully propagated to layers below as shown in Fig.~\ref{fig:quasilinear_halfadders}.
By not adding these empty bits using superfluous half-adders we avoid inflation of the overall Hamiltonian which would be due to each half-adder introducing new high-order terms.

The half-adder circuit results in a sum string $s$ that contains the number of turns the protein has taken in the $\pm k$ direction between any two residues.
In our notation, it holds that,
\begin{equation}
	s^r_{\pm k}(i,j) = r^{th}\,\text{digit of} \sum^{j-1}_{p=i} d^p_{\pm k}.
\end{equation}

\textbf{Analysis of the circuit complexity.}
Using the technique outlined in the previous subsection, we obtained a significant quadratic to quasilinear improvement in circuit complexity for sum strings, which are basic building blocks for the subcomponents of the Hamiltonian, therefore, propagating through all subsequent expressions.
This section derives and proves this bound.

The total number of half-adders $h_\text{total}(n)$ in the sum string circuit of $n$ bits is (counting half-adders from bottom up),
\begin{equation}
    h_\text{total} = 1 + 2 + \cdots + n = \frac{n(n+1)}{2}.
\end{equation}

The circuit complexity in terms of number of half-adders involved is then $O(n^2)$.
However, since the sum of $n$ bits only needs $\lceil \log_2 (n+1) \rceil$ output bits, the number of bits $n_\text{redun}$ not containing any information output is,
\begin{equation}
    n_\text{redun} = n - \lceil \log_2 (n+1) \rceil.
\end{equation}

The number of associated half-adders that cannot propagate any information to output bits is equal to $h_\text{total}(n_\text{redun})$, by isomorphism to the sum string circuit of $n_\text{redun}$ binary variables (see Fig.~\ref{fig:quasilinear_halfadders}).

The necessary number of half-adders $h_\text{improv}$ in the improved circuit for addition of $n$ binary variables is then,
\begin{align}
    &h_\text{improv}(n) = h_\text{total}(n) - h_\text{total}(n_\text{redun}) \nonumber \\
    =&\,\,\, \frac{n^2 + n}{2} - \frac{n^2_\text{redun}+n_\text{redun}}{2} \nonumber \\
    =&\,\,\, \frac{n^2 + n}{2} -\frac{(n - \lceil \log_2 (n+1) \rceil)^2+n - \lceil \log_2 (n+1) \rceil}{2} \nonumber \\
    =&\,\,\, \frac{2n\lceil \log_2 (n+1) \rceil - \lceil \log_2 (n+1) \rceil ^2 + \lceil \log_2 (n+1) \rceil}{2}.
\end{align}

It follows that $h_\text{improv}(n) \in O(n\log n)$, hence, providing a significant decrease in the circuit complexity, from quadratic to quasilinear.

\pgfdeclarelayer{background} 
\pgfsetlayers{background,main} 

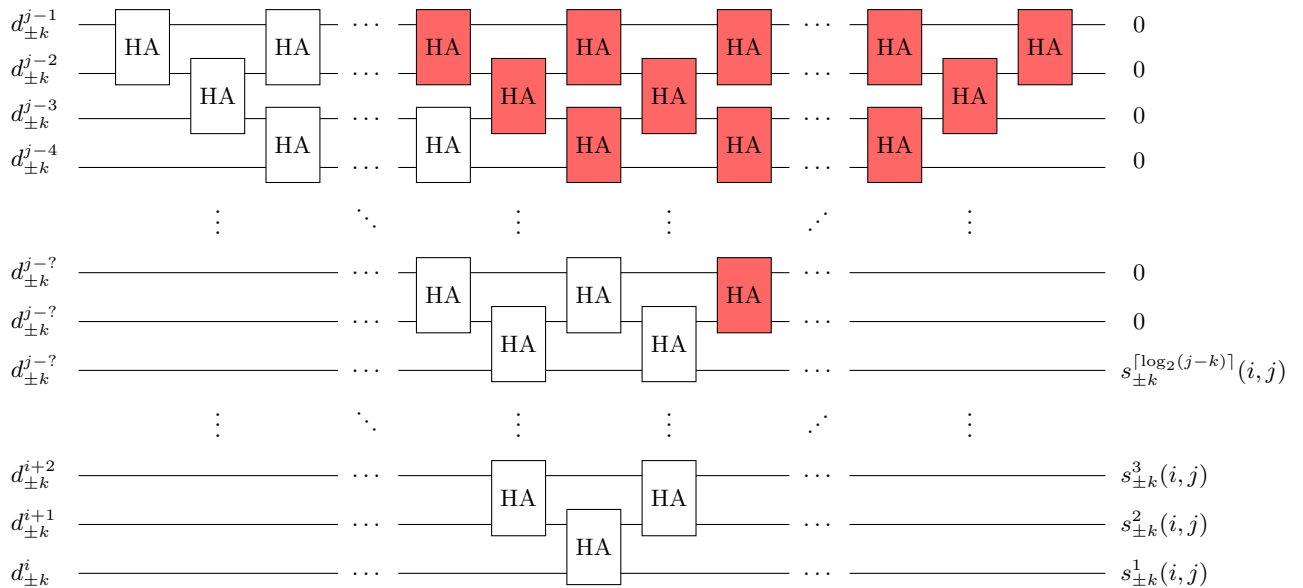
\begin{figure*}
\begin{tikzpicture}

\begin{pgfonlayer}{main}
  \node[draw, minimum width=0.5cm, minimum height=1cm, text width=0.5cm, fill=white, align=center] (HA1) at (0,4) {HA};
  \node[draw, minimum width=0.5cm, minimum height=1cm, text width=0.5cm, fill=white, align=center] (HA2) at (2,4) {HA};
  \node[draw, minimum width=0.5cm, minimum height=1cm, text width=0.5cm, fill=red!60, align=center] (HA3) at (4,4) {HA};
  \node[draw, minimum width=0.5cm, minimum height=1cm, text width=0.5cm, fill=red!60, align=center] (HA4) at (6,4) {HA};
  \node[draw, minimum width=0.5cm, minimum height=1cm, text width=0.5cm, fill=red!60, align=center] (HA5) at (8,4) {HA};
  \node[draw, minimum width=0.5cm, minimum height=1cm, text width=0.5cm, fill=red!60, align=center] (HA5) at (10,4) {HA};
  \node[draw, minimum width=0.5cm, minimum height=1cm, text width=0.5cm, fill=red!60, align=center] (HA5) at (12,4) {HA};

\node[draw, minimum width=0.5cm, minimum height=1cm, text width=0.5cm, fill=white, align=center] (HA6) at (1,3.35) {HA};
  \node[draw, minimum width=0.5cm, minimum height=1cm, text width=0.5cm, fill=red!60, align=center] (HA7) at (5,3.35) {HA};
  \node[draw, minimum width=0.5cm, minimum height=1cm, text width=0.5cm, fill=red!60, align=center] (HA8) at (7,3.35) {HA};
  \node[draw, minimum width=0.5cm, minimum height=1cm, text width=0.5cm, fill=red!60, align=center] (HA8) at (11,3.35) {HA};

  \node[draw, minimum width=0.5cm, minimum height=1cm, text width=0.5cm, fill=white, align=center] (HA9) at (2,2.7) {HA};
  \node[draw, minimum width=0.5cm, minimum height=1cm, text width=0.5cm, fill=white, align=center] (HA10) at (4,2.7) {HA};
  \node[draw, minimum width=0.5cm, minimum height=1cm, text width=0.5cm, fill=red!60, align=center] (HA11) at (6,2.7) {HA};
  \node[draw, minimum width=0.5cm, minimum height=1cm, text width=0.5cm, fill=red!60, align=center] (HA11) at (8,2.7) {HA};
  \node[draw, minimum width=0.5cm, minimum height=1cm, text width=0.5cm, fill=red!60, align=center] (HA11) at (10,2.7) {HA};

\node[draw, minimum width=0.5cm, minimum height=1cm, text width=0.5cm, fill=white, align=center] (HA10) at (4,0.7) {HA};
  \node[draw, minimum width=0.5cm, minimum height=1cm, text width=0.5cm, fill=white, align=center] (HA11) at (6,0.7) {HA};
  \node[draw, minimum width=0.5cm, minimum height=1cm, text width=0.5cm, fill=red!60, align=center] (HA11) at (8,0.7) {HA};

  \node[draw, minimum width=0.5cm, minimum height=1cm, text width=0.5cm, fill=white, align=center] (HA12) at (7,0.05) {HA};
  \node[draw, minimum width=0.5cm, minimum height=1cm, text width=0.5cm, fill=white, align=center] (HA13) at (5,0.05) {HA};

  \node[draw, minimum width=0.5cm, minimum height=1cm, text width=0.5cm, fill=white, align=center] (HA12) at (7,-2) {HA};
  \node[draw, minimum width=0.5cm, minimum height=1cm, text width=0.5cm, fill=white, align=center] (HA13) at (5,-2) {HA};

	\node[draw, minimum width=0.5cm, minimum height=1cm, text width=0.5cm, fill=white, align=center] (HA14) at (6,-2.65) {HA};

\end{pgfonlayer}

\begin{pgfonlayer}{background}


  \node[minimum width=0.5cm, minimum height=1cm, text width=0.5cm, align=center] (HA1) at (-1.5,4.3) {$d^{j-1}_{\pm k}$};
  \node[minimum width=0.5cm, minimum height=1cm, text width=0.5cm, align=center] (HA1) at (-1.5,3.7) {$d^{j-2}_{\pm k}$};
  \node[minimum width=0.5cm, minimum height=1cm, text width=0.5cm, align=center] (HA1) at (-1.5,3.1) {$d^{j-3}_{\pm k}$};
  \node[minimum width=0.5cm, minimum height=1cm, text width=0.5cm, align=center] (HA1) at (-1.5,2.5) {$d^{j-4}_{\pm k}$};

  \node[minimum width=0.5cm, minimum height=1cm, text width=0.5cm, align=center] (HA1) at (-1.5,1.0) {$d^{j-?}_{\pm k}$};
  \node[minimum width=0.5cm, minimum height=1cm, text width=0.5cm, align=center] (HA1) at (-1.5,0.35) {$d^{j-?}_{\pm k}$};
  \node[minimum width=0.5cm, minimum height=1cm, text width=0.5cm, align=center] (HA1) at (-1.5,-0.3) {$d^{j-?}_{\pm k}$};

  \node[minimum width=0.5cm, minimum height=1cm, text width=0.5cm, align=center] (HA1) at (-1.5,-1.7) {$d^{i+2}_{\pm k}$};
  \node[minimum width=0.5cm, minimum height=1cm, text width=0.5cm, align=center] (HA1) at (-1.5,-2.35) {$d^{i+1}_{\pm k}$};
  \node[minimum width=0.5cm, minimum height=1cm, text width=0.5cm, align=center] (HA1) at (-1.5,-3) {$d^{i}_{\pm k}$};

  \node[minimum width=0.5cm, minimum height=1cm, text width=0.5cm, align=center] (HA1) at (13.25,4.3) {0};
  \node[minimum width=0.5cm, minimum height=1cm, text width=0.5cm, align=center] (HA1) at (13.25,3.7) {0};
  \node[minimum width=0.5cm, minimum height=1cm, text width=0.5cm, align=center] (HA1) at (13.25,3.1) {0};
  \node[minimum width=0.5cm, minimum height=1cm, text width=0.5cm, align=center] (HA1) at (13.25,2.5) {0};

  \node[minimum width=0.5cm, minimum height=1cm, text width=0.5cm, align=center] (HA1) at (13.25,1.0) {0};
  \node[minimum width=0.5cm, minimum height=1cm, text width=0.5cm, align=center] (HA1) at (13.25,0.35) {0};
  \node[minimum width=0.5cm, minimum height=1cm, text width=0.5cm, align=center] (HA1) at (13.25,-0.3) {$s^{\lceil\log_2(j-k)\rceil}_{\pm k}(i,j)$};

  \node[minimum width=0.5cm, minimum height=1cm, text width=0.5cm, align=center] (HA1) at (13.25,-1.7) {$s^{3}_{\pm k}(i,j)$};
  \node[minimum width=0.5cm, minimum height=1cm, text width=0.5cm, align=center] (HA1) at (13.25,-2.35) {$s^{2}_{\pm k}(i,j)$};
  \node[minimum width=0.5cm, minimum height=1cm, text width=0.5cm, align=center] (HA1) at (13.25,-3) {$s^{1}_{\pm k}(i,j)$};

  \draw[-] (-0.85,4.3) -- (2.6,4.3);
  \draw[-] (-0.85,3.65) -- (2.6,3.65);

  \node[minimum width=0.5cm, minimum height=1cm, text width=0.5cm, fill=white, align=center] (test) at (3.0,4.3) {\ldots};
  \node[minimum width=0.5cm, minimum height=1cm, text width=0.5cm, fill=white, align=center] (test) at (3.0,3.65) {\ldots};

  \draw[-] (3.4,4.3) -- (8.6,4.3);
  \draw[-] (3.4,3.65) -- (8.6,3.65);

  \node[minimum width=0.5cm, minimum height=1cm, text width=0.5cm, fill=white, align=center] (test) at (9,4.3) {\ldots};
  \node[minimum width=0.5cm, minimum height=1cm, text width=0.5cm, fill=white, align=center] (test) at (9,3.65) {\ldots};

  \draw[-] (9.4,4.3) -- (12.8,4.3);
  \draw[-] (9.4,3.65) -- (12.8,3.65);

  \draw[-] (-0.85,3.05) -- (2.6,3.05);
  \draw[-] (-0.85,2.4) -- (2.6,2.4);

  \node[minimum width=0.5cm, minimum height=1cm, text width=0.5cm, fill=white, align=center] (test) at (3.0,3.05) {\ldots};
  \node[minimum width=0.5cm, minimum height=1cm, text width=0.5cm, fill=white, align=center] (test) at (3.0,2.4) {\ldots};

  \draw[-] (3.4,3.05) -- (8.6,3.05);
  \draw[-] (3.4,2.4) -- (8.6,2.4);

  \node[minimum width=0.5cm, minimum height=1cm, text width=0.5cm, fill=white, align=center] (test) at (9.0,3.05) {\ldots};
  \node[minimum width=0.5cm, minimum height=1cm, text width=0.5cm, fill=white, align=center] (test) at (9.0,2.4) {\ldots};

  \draw[-] (9.4,3.05) -- (12.8,3.05);
  \draw[-] (9.4,2.4) -- (12.8,2.4);

  \draw[-] (-0.85,1.0) -- (2.6,1.0);
  \draw[-] (-0.85,0.35) -- (2.6,0.35);

  \node[minimum width=0.5cm, minimum height=1cm, text width=0.5cm, fill=white, align=center] (test) at (3.0,1) {\ldots};
  \node[minimum width=0.5cm, minimum height=1cm, text width=0.5cm, fill=white, align=center] (test) at (3.0,0.35) {\ldots};

  \draw[-] (3.4,1.0) -- (8.6,1.0);
  \draw[-] (3.4,0.35) -- (8.6,0.35);

  \node[minimum width=0.5cm, minimum height=1cm, text width=0.5cm, fill=white, align=center] (test) at (9.0,1) {\ldots};
  \node[minimum width=0.5cm, minimum height=1cm, text width=0.5cm, fill=white, align=center] (test) at (9.0,0.35) {\ldots};

  \draw[-] (9.4,1.0) -- (12.8,1.0);
  \draw[-] (9.4,0.35) -- (12.8,0.35);

  \draw[-] (-0.85,-0.3) -- (2.6,-0.3);

  \node[minimum width=0.5cm, minimum height=1cm, text width=0.5cm, fill=white, align=center] (test) at (3.0,-0.3) {\ldots};

  \draw[-] (3.4,-0.3) -- (8.6,-0.3);

  \node[minimum width=0.5cm, minimum height=1cm, text width=0.5cm, fill=white, align=center] (test) at (9.0,-0.3) {\ldots};

  \draw[-] (9.4,-0.3) -- (12.8,-0.3);

  \draw[-] (-0.85,-1.7) -- (2.6,-1.7);
  \draw[-] (-0.85,-2.35) -- (2.6,-2.35);
  \draw[-] (-0.85,-3) -- (2.6,-3);

  \node[minimum width=0.5cm, minimum height=1cm, text width=0.5cm, fill=white, align=center] (test) at (3.0,-1.7) {\ldots};
  \node[minimum width=0.5cm, minimum height=1cm, text width=0.5cm, fill=white, align=center] (test) at (3.0,-2.35) {\ldots};
  \node[minimum width=0.5cm, minimum height=1cm, text width=0.5cm, fill=white, align=center] (test) at (3.0,-3) {\ldots};

  \draw[-] (3.4,-1.7) -- (8.6,-1.7);
  \draw[-] (3.4,-2.35) -- (8.6,-2.35);
  \draw[-] (3.4,-3) -- (8.6,-3);

  \node[minimum width=0.5cm, minimum height=1cm, text width=0.5cm, fill=white, align=center] (test) at (9.0,-1.7) {\ldots};
  \node[minimum width=0.5cm, minimum height=1cm, text width=0.5cm, fill=white, align=center] (test) at (9.0,-2.35) {\ldots};
  \node[minimum width=0.5cm, minimum height=1cm, text width=0.5cm, fill=white, align=center] (test) at (9.0,-3) {\ldots};

  \draw[-] (9.4,-1.7) -- (12.8,-1.7);
  \draw[-] (9.4,-2.35) -- (12.8,-2.35);
  \draw[-] (9.4,-3) -- (12.8,-3);

  \node[rotate=90, minimum width=0.5cm, minimum height=1cm, text width=0.5cm, fill=white, align=center] (test) at (1.0,1.7) {\ldots};
  \node[rotate=90, minimum width=0.5cm, minimum height=1cm, text width=0.5cm, fill=white, align=center] (test) at (1.0,-1.0) {\ldots};

  \node[rotate=90, minimum width=0.5cm, minimum height=1cm, text width=0.5cm, fill=white, align=center] (test) at (5.0,1.7) {\ldots};
  \node[rotate=90, minimum width=0.5cm, minimum height=1cm, text width=0.5cm, fill=white, align=center] (test) at (7.0,1.7) {\ldots};
  \node[rotate=90, minimum width=0.5cm, minimum height=1cm, text width=0.5cm, fill=white, align=center] (test) at (5.0,-1.0) {\ldots};
  \node[rotate=90, minimum width=0.5cm, minimum height=1cm, text width=0.5cm, fill=white, align=center] (test) at (7.0,-1.0) {\ldots};

  \node[rotate=90, minimum width=0.5cm, minimum height=1cm, text width=0.5cm, fill=white, align=center] (test) at (11.0,1.7) {\ldots};
  \node[rotate=90, minimum width=0.5cm, minimum height=1cm, text width=0.5cm, fill=white, align=center] (test) at (11.0,-1.0) {\ldots};

  \node[rotate=-45, minimum width=0.5cm, minimum height=1cm, text width=0.5cm, fill=white, align=center] (test) at (3.0,1.7) {\ldots};
  \node[rotate=45, minimum width=0.5cm, minimum height=1cm, text width=0.5cm, fill=white, align=center] (test) at (9.0,1.7) {\ldots};
  \node[rotate=-45, minimum width=0.5cm, minimum height=1cm, text width=0.5cm, fill=white, align=center] (test) at (3.0,-1) {\ldots};
  \node[rotate=45, minimum width=0.5cm, minimum height=1cm, text width=0.5cm, fill=white, align=center] (test) at (9.0,-1) {\ldots};

\end{pgfonlayer}


\end{tikzpicture}
\caption{\label{fig:quasilinear_halfadders} Circuit diagram demonstrating the quadratic to quasilinear improvement in circuit complexity outlined in this paper.
Most output bits on the right side are zeros since the sum of $(j-k)$ input bits only requires $\lceil \log_2(j-k) \rceil$ output bits.
Therefore, all half-adders coloured in red are superfluous and can be omitted, which reduces the number of half-adders from quadratic to quasilinear.}
\end{figure*}

\textbf{Construction of $H_\text{olap}$.} With the construction of the sum strings we have already done the majority of the work needed for the construction of $H_\text{olap}$.
The sum strings tell us the exact number of turns that the protein chain has taken into the $\pm k$ direction.
Thus, two residues overlap if and only if all corresponding $\pm k$ sum strings sum to zero.
We can test if two sum bits are different using an XNOR function,
\begin{equation}
	\text{XNOR}(p,q) = 1-p-q+2pq.
\end{equation}

To test if two residues overlap we compare each bit in the two sum strings using the XNOR function and take the product over all the respective XNOR results and over all dimensions: 
\begin{equation}
	\label{equ:h_olap_ij}
    H_\text{olap}(i,j) = \prod^3_{k=1} \Big( \prod^{\lceil\log_2(j-i)\rceil}_{r=1}\text{XNOR}(s^r_{+k}(i,j),s^r_{-k}(i,j)) \Big).
\end{equation}

From the definition of the XNOR function it follows that $H_\text{olap}(i, j) = 1$ if and only if amino acid pair $(i, j)$ overlaps.
To penalize any overlap, we take the double summation over all possible amino acid pairs that could possibly overlap and compare their sum strings using Eq.~\ref{equ:h_olap_ij},
\begin{equation}
	H_\text{olap} = \lambda_\text{olap} \sum^{N-2}_{i=1} \sum^{\lfloor(N-i)/2\rfloor}_{j=1} H_\text{olap}(i,i+2j) .
\end{equation}

\textbf{Construction of $H_\text{pair}$.} To construct this subcomponent we again leverage the previously defined sum strings (see Section \ref{sec:turn_circuit}).
Two residues interact with each other, that is have lattice distance of 1, if and only if exactly one $\pm k$ (direction) pair of sum strings sum to 1 and both of the other direction sum strings pairs sum exactly to 0.
The following adjacency function \cite{babbush2012construction} captures the condition above:
\begin{align}
	a_k(i,j) = &\Bigg[ \prod_{w \neq k} \Big( \prod^{\lceil \log_2(j-i)\rceil}_{r=1} \text{XNOR}(s^r_{+w}(i,j),s^r_{-w}(i,j)) \Big) \Bigg] \nonumber \\
			   &*\Bigg[ \text{XOR} (s^1_{+k}(i,j),s^1_{-k}(i,j)) \nonumber \\
			   &*\prod^{\lceil \log_2(j-i)\rceil}_{r=2} \text{XNOR} (s^r_{+k}(i,j),s^r_{-k}(i,j)) \nonumber \\
			   &+ \sum^{\lceil \log_2(j-i)\rceil}_{p=2} \Big( \text{XOR} (s^{p-1}_{+k}(i,j),s^p_{+k}(i,j)) \nonumber \\
			   &*\prod^{p-2}_{r=1} \text{XNOR} (s^r_{+k}(i,j),s^{r+1}_{+k}(i,j)) \nonumber \\
			   &*\prod^{p}_{r=1} \text{XOR} (s^r_{+k}(i,j),s^{r}_{-k}(i,j)) \nonumber \\
			   &*\prod^{\lceil \log_2(j-i)\rceil}_{r=p+1} \text{XNOR} (s^r_{+k}(i,j),s^{r}_{-k}(i,j)) \Big) \Bigg].
\end{align}

This function evaluates to $1$ if two sum strings differ by exactly $1$ and $0$ otherwise.
Thus, to determine if two amino acids interact we need to sum the adjacency function over all three dimensions for this amino acid pair,
\begin{equation}
	H_\text{pair}(i,j) = P_{ij}\sum^{3}_{k=1} a_k(i,j),
\end{equation}
where $P_{ij}$ is again the interaction matrix obtained using either the HP or the MJ potential.
The expression is multiplied with the $i,j$-th matrix element to favour residue interaction by effectively lowering the overall energy of the problem Hamiltonian if two strongly interacting amino acids are adjacent on the lattice.

Finally, we sum over all amino acid pairs that can possibly interact with each other:
\begin{equation}
	H_\text{pair} = \sum^{n-3}_{i=1}  \sum^{(N-i-1)/2}_{j=1} H_\text{pair}(i,1+i+2j) .
\end{equation}

\subsection{\label{sec:binary_flag_encoding}{Binary flag based encoding}}

As an alternative approach to encoding coordinates of the vertices of the lattice protein fold, or encoding its edges we can populate the lattice grid space with labels, each associated with every possible position of the given amino acid.
While this requires an exponential number of qubits, the size of the lattice space can be bounded by the radius, hence, providing a linear mapping in terms of the number of qubits, as implemented in Section \ref{sec:nested_shell_encoding}.

\subsubsection{\label{sec:nested_shell_encoding}Nested shell encoding}

The nested shell encoding is based on two fundamental ideas.
First, it enforces an arbitrary, pre-selected bound on the radius of the cubic lattice space we allow the lattice fold to explore.
Second, it does not use qubits to directly encode the information about the position of the amino acids, but rather uses them as binary flags signalling the presence of the given amino acid at a particular position in a cubic grid.
These ideas were first used by \citet{babbush2012construction} to formulate the \textit{diamond encoding}, which was designed for planar lattices.

Although using fairly large number of qubits to represent each amino acid, this encoding has the advantage of being 2-local and quite sparse, hence, avoiding two major hidden costs of the previous encodings, namely the ancilla introduction due to conversion from k-local to 2-local Hamiltonian and it significantly reduces the overhead in the minor embedding to the hardware Chimera architecture of the quantum processing unit.

First we define necessary concepts in order to introduce the nested shell encoding, which partitions the cubic grid space into a set of nested vertex shells.
Consider a cubic grid graph $G=(V, E)$ with a point of origin $O \in V$.
We define \textit{nested shell} $S_0$ as $S_0={O}$ and nested shell $S_i, i>0$ as,
\begin{equation}
    S_i = \{v\,|\, v \text{ is adjacent to } w, w \in S_{i-1}\}.
\end{equation}

Let $p$ be a lattice protein of length $n$.
Next we define the \textit{amino acid vertex set} $V_i$ as the union of all nested shells that amino acid on the $i$-th position can occupy, $i \leq n$.
\begin{equation}
    V_i = \bigcup_{j\in J} S_j,
\end{equation}
where,
\begin{equation}
	J = \begin{cases}
        \{i\} & \text{if } i\leq 2,\\
        \{1,3,\ldots, i\} & \text{if } i > 2 \text{ and i is odd} , \\
        \{2,4,\ldots, i\} & \text{if } i > 2 \text{ and i is even}.
\\
	\end{cases}
\end{equation}

Each vertex in the vertex set $V_i$ represents a possible position of the $i$-th amino acid and hence, it gets assigned one binary qubit flag.
We define the \textit{amino acid qubitset} $Q_i$, set of all flags corresponding to the $i$-th amino acid as,
\begin{equation}
    Q_i = \{q_{\gamma(i)}, \ldots, q_{\gamma(i+1)-1}\},
\end{equation}
where,
\begin{equation}
    \gamma(i) = \sum_{j=0}^{i-1} |V_j|.
\end{equation}

With the definitions above in place, we can now proceed to the derivation of the Hamiltonian for the nested shell encoding:
\begin{equation}
    H = H_\text{one} + H_\text{conn} + H_\text{pair} + H_\text{olap}
\end{equation}

Each subcomponent of the Hamiltonian serves a critical function.
$H_\text{one}$ ensures that no two qubits from the same amino acid qubitset are signalling at the same time.
Given a lattice protein of length $n$, it holds that,
\begin{equation}
    H_\text{one} = \lambda_\text{one} \sum^{n-1}_{i=0} \sum_{q_a, q_b \in Q_i, a<b} q_a q_b.
\end{equation}

To ensure that the protein is connected, we need to favour adjacent grid positions for qubit sets of subsequent amino acids.
This is the purpose of the $H_\text{conn}$:
\begin{equation}
    H_\text{conn} = \lambda_\text{conn} \Big( n - 1 - \sum_{i=0}^{n-1} \sum_{q_d\in Q_i} \sum_{q_u\in\eta(q) \cap Q_{i+1}} q_d q_u \Big).
\end{equation}
where $\eta(q)$ denotes the set of all qubits assigned to a vertex adjacent to the vertex assigned to qubit $q$.

The $H_\text{olap}$ part of the Hamiltonian introduces the constraints necessary to prevent two qubits from different amino acids claiming their assigned vertex as occupied by their amino acid at the same time.
Let $\theta(v)$ be a set of all qubits occupying a given grid vertex $v \in V$ of the lattice grid graph $G=(V, E)$.
Then $H_\text{olap}$ is given as
\begin{equation}
    H_\text{olap} = \lambda_\text{olap} \sum_{v\in V} \sum_{q_a, q_b \in \theta(v), a<b} q_a q_b
\end{equation}

As in the previous encodings, to properly assess the interaction between the $i$-th and $j$-th amino acid we construct the interaction matrix $P$ that contains interaction strengths between all amino acids that can possibly interact on the given cubic lattice grid, restricted by parity and distance (amino acids vertices need to be at least 3 edges apart to interact).
Let $\omega$ be a mapping from a qubit to the position of its amino acid in the sequence.
Then $H_\text{pair}$ is defined as,
\begin{equation}
    H_\text{pair} = \frac{1}{2} \sum_{i=0}^{n-1} \sum_{q_a \in Q_i} \sum_{q_b \in \eta(q_a} P_{\omega(q_a),\omega(q_b)} q_a q_b.
\end{equation}

%
%

\section{Results \& Discussion}
\label{sec:results_discussion}
\begin{figure}[h!]
	\centering \includegraphics[width=0.5\textwidth]{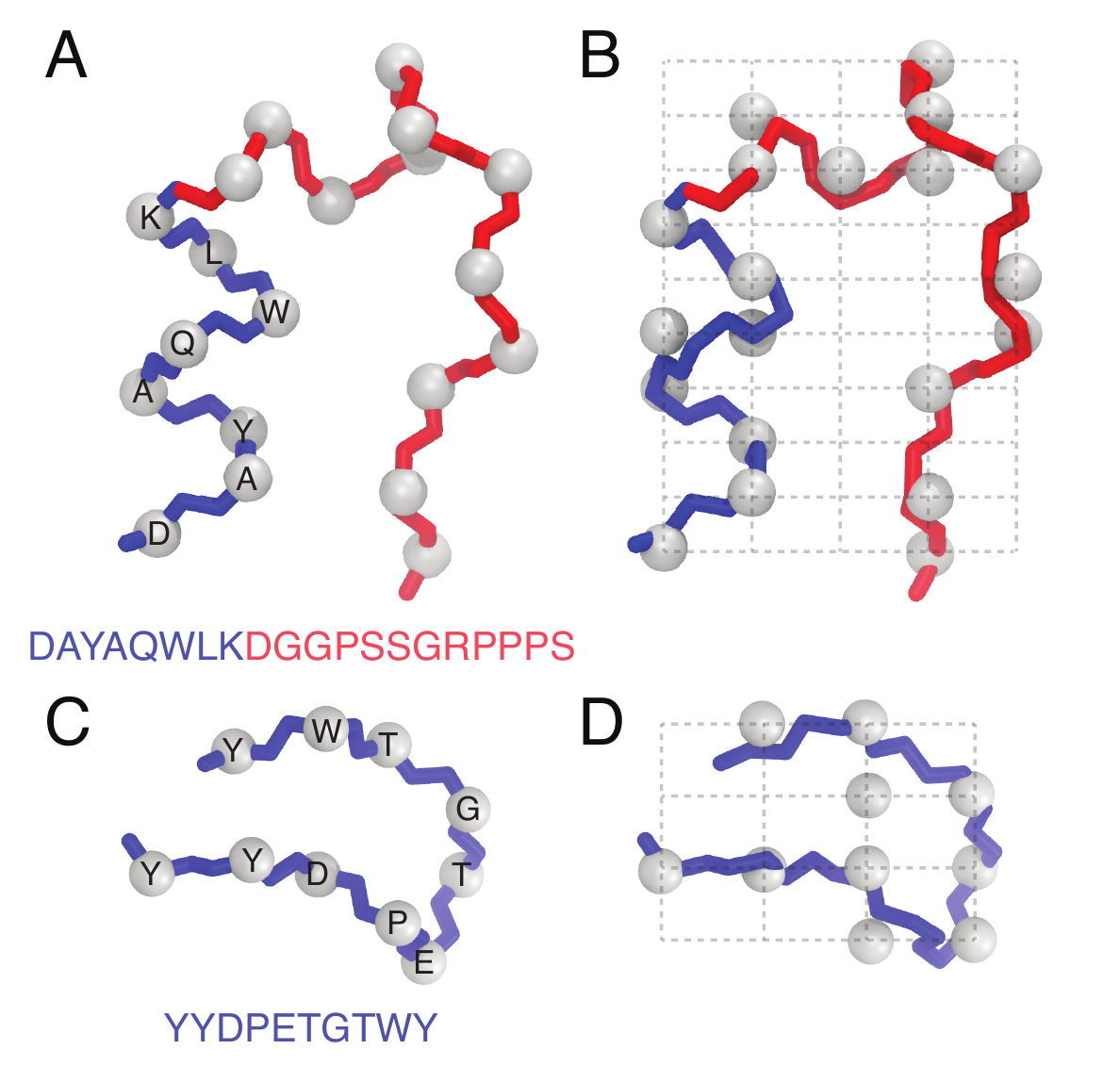}
	\caption{\label{fig:real_proteins} Molecular representation of Trp-Cage and Chignolin. On the left, the NMR structures of Trp-Cage (Fig. A, PDB: 2JOF) and Chignolin (Fig. C, PDB: 2RVD) are shown. In both representations, only backbone atoms are shown and C-alpha atoms are highlighted as white spheres. On the right, Trp-Cage (Fig. B) and Chignolin (Fig. D) are shown with C-alpha atoms discretized onto a cubic lattice, where CA-CA distance was fixed to be 3.8 Angstroms.}
\end{figure}

For our experimental implementations, we chose two well-studied proteins, Chignolin for the planar lattice and Trp-Cage for the cubic lattice.
Fig.~\ref{fig:real_proteins}A and \ref{fig:real_proteins}C show the molecular structures, solved using experimentally provided contact restraints from nuclear magnetic resonance (NMR) of the Trp-Cage fragment (blue) and Chignolin, respectively.
First, we will discuss the results for folding Chignolin with 10 residues on a planar square lattice using the D-Wave 2000Q quantum processing unit.

In previous work, \citet{perdomo2012finding} achieved folding of a protein with 6 residues (PSVKMA) on a planar lattice.
Chignolin is a well-studied artificial mini-protein with 10 residues that has been shown to fold into a $\beta$-hairpin structure in water \citep{honda200410}.
The advances in quantum hardware alone would have not sufficed to fold lattice proteins of this size, hence, our algorithmic improvements were crucial in the ability to surpass the state-of-the-art.
To perform the experiments, we used D-Wave's newest chip generation 2000Q with 2048 superconducting qubits arranged in a Chimera graph with sparse connectivity \citep{yang2016graph}.

For the folding of Chignolin we generated the Hamiltonian with the turn ancilla encoding locally on our server.
We then split the large Hamiltonian into $1024$ subproblems and requested ten batches of $10000$ samples each from the quantum processor.
We made use of spin reversal transforms in order to increase the quality of our solutions and we additionally performed single-flip gradient descent to drive solutions of Hamming distance 1 to the ground state.
For verification purposes we used a straightforward Monte-Carlo solver on classical hardware to find the correct ground state of Chignolin.

The most important measure to evaluate stochastic algorithms is the time-to-solution (TTS) metric.
In most studies of this kind, the number of repetitions $R_{99}$ required in order to obtain the solution at least once with $99\%$ certainty is computed in order to evaluate the TTS.
The expression for $R_{99}$ is given by,
\begin{equation}
	R_{99} = \lceil \frac{\log(1-0.99)}{\log(1-p_s)}  \rceil,
\end{equation}
where $p_s$ is the success probability of obtaining the correct solution with only one execution of the solver.

\begin{figure}[h!]
	\centering \includegraphics[width=0.5\textwidth]{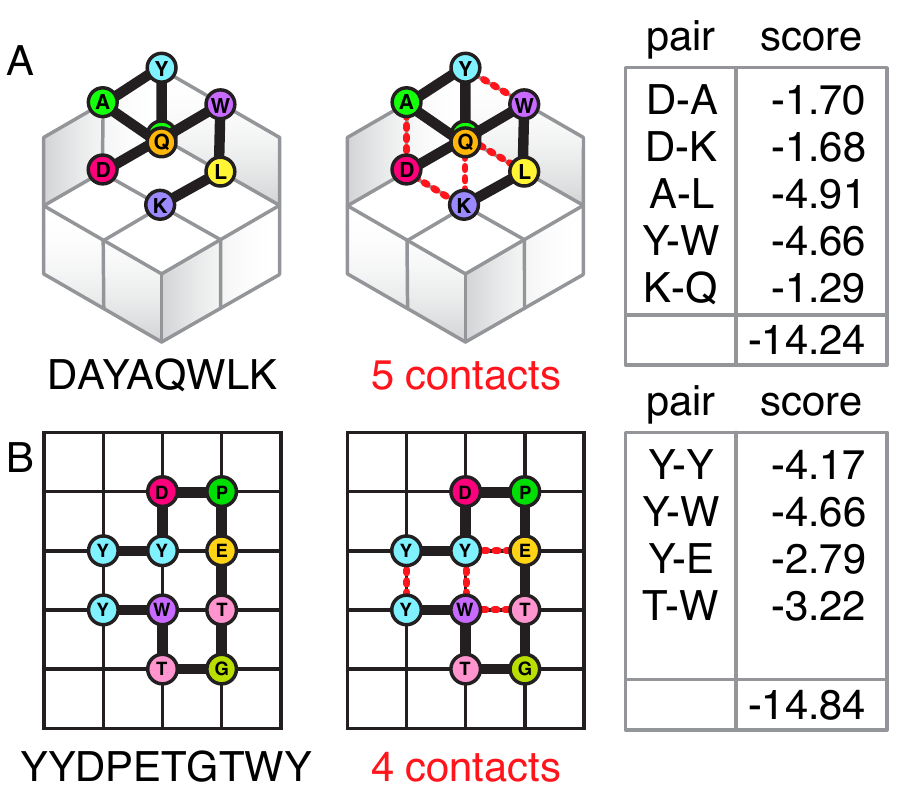}
	\caption{\label{fig:q_lattice_results} Visualization of the ground state lattice folds for A) the 8 residue Trp-Cage snippet (DAYAQWLK) on a cubic lattice and B) the 10 residue Chignolin (YYDPETGTWY) on a square lattice.
Both lattice folds were obtained with the D-Wave 2000Q quantum annealer.
The tables on the right show the MJ interaction strengths for the interacting amino acid pairs in the depicted lattice folds and the obtained total ground state energy.}
\end{figure}

The correct Chignolin lattice fold and ground state energy that we obtained from the QPU are shown in Fig.~\ref{fig:q_lattice_results}.
In our experiment, we collected a total of 102 400 000 samples from the quantum processing unit (QPU) of which 24950 samples were found to be the correct ground state.
This implies a success probability of $p_s = 0.000244$.
However, note that only one out of the $2^{10}$ subproblems actually included the correct solution in its solution space.
For each subproblem we requested 100.000 samples which means that within this subproblem we had a success probability of $p_s^{sub} = 0.2495$.
It is, of course, impossible to know beforehand which subproblem will contain the correct solution and, thus, we will only use $p_s$ for further analysis.
Folding Chignolin on the 2D lattice in this experimental setup would require $R_{99} = 190262$ samples to be drawn in order to obtain the correct solution at least once.
We used an annealing schedule with an annealing time of $t_{sample}=20\mu s$ which leads to a time-to-solution of $TTS = 0.377s$.

Fig.~\ref{fig:real_proteins}D shows the experimentally determined structure of Chignolin with its C-alpha atoms discretized onto a cubic lattice, where CA-CA distance was fixed to be 3.8 Angstroms.
This discretization was performed by doing a global optimization of CA positions to the grid using the LatFit algorithm \citep{mann2012producing}.
Specifically, this algorithm minimizes the distance RMSD between the original and the produced lattice protein.
The C-alpha atom RMSD between the discretized and real space positions is 0.85 Angstroms only for Chignolin.
It is particularly interesting to examine the similarity between the square lattice fold from the QPU in Fig.~\ref{fig:q_lattice_results}B and the discretized version of the real structure in Fig.~\ref{fig:real_proteins}D.
Most noticeable is the contact between the two Y residues at the two ends and the next pair between a Y and a W residue of the protein.
It is exactly this kind of contact information that can be retrieved from lattice folds and subsequently be used to speed up classical molecular dynamics simulations.

For the 3D lattice folding experiment, we used an 8 amino acid snippet of Trp-Cage, another well-studied mini-protein with interesting secondary structure such as an $\alpha$-helical component \citep{neidigh2002designing}.
For this amino acid sequence (DAYAQWLK) we again generated the Hamiltonian using the turn ancilla encoding locally on our server.
However, this time we split the large Hamiltonian into $2^{12}$ subproblems and requested only five batches of $10000$ readouts each from the quantum processor via cloud access.
Again for verification purposes, the correct lattice fold was obtained with a classical Monte-Carlo solver on classical hardware.

The 3D lattice fold and ground state energy we obtained for this protein are shown in Fig.~\ref{fig:q_lattice_results}A.
It is important to highlight, that this is the first time that a three dimensional lattice protein has been folded on quantum computing hardware.
Interestingly, the obtained ground state lattice fold looks a lot like the beginning of an $\alpha$-helix which matches up with the real structure.
Fig.~\ref{fig:real_proteins}B shows the Trp-Cage structure with C-alpha atoms discretized onto a cubic lattice, where the CA-CA distance was again fixed to be 3.8 Angstroms.
This discretization was also performed using the LatFit algorithm \citep{mann2012producing}.
Specifically, this algorithm minimizes the distance RMSD between the original and the produced lattice protein.
The C-alpha atom RMSD between the discretized and real space positions is only 0.83 Angstroms for Trp-Cage.
This RMSD value suggests that the cubic lattice discretization error is small enough such that moderately accurate atomistic models may be reconstructed from single-point residue lattice folds.
Comparing the 3D lattice fold from the QPU and the actual structure shows strong similarity since we can clearly see the first D residue interacting with the second A residue which are exactly one helix turn away from each other.
Additionally, the Q and K residue interaction can clearly be seen in the experimentally determined structure.

For the Trp-Cage fragment we requested a total of 204 800 000 samples from the QPU of which 4957 samples represented the correct lattice fold.
Therefore, the success probability within the subproblem that contained the correct solution is $p_s^{sub}=0.099$.
The overall success probability is $p_s=2.42\times10^{-5}$ and, thus, $R_{99}=190262$ samples are needed in order to have 99\% confidence that we obtained the solution at least once.
Using an annealing time of $t_{sample}=20\mu s$ results in a time-to-solution of $TTS=3.805s$.

In order to improve the TTS in both experiments we could further decrease the number of subproblems since for now we only used roughly 200 qubits of the quantum processor for each run in order to enable fast embeddings onto the Chimera graph.
In the near future, we will try to push the limits by decreasing the number of subproblems and maximizing the use of the hardware.
Another possibility for improvement is better postprocessing of solutions such as implementing multi spin flips instead of single spin flips in the postprocessing step.
Even though currently not practically implementable, it should be noted that our current folding algorithm is parallelizable over multiple QPUs since the subproblems do not need to be executed sequentially.




\section{Conclusions and Future Work}
\label{sec:conclusion}

Lattice models are powerful tools to investigate the fundamental principles of protein folding.
The representation of proteins on a discrete lattice enables computationally rigorous investigation of protein sequence and structure relationships.
In this work, we describe advances to lattice folding algorithms on quantum annealing devices that may extend the capability of these models to challenging problems in structural biology and drug design that are currently computationally intractable.
Specifically, we generalize the work of \citet{babbush2012construction} and \citet{perdomo2012finding} to enable the folding of proteins on a cubic lattice, and demonstrate significant improvements in circuit complexity in the turn circuit encoding.
Advances in quantum hardware and our algorithmic improvements have enabled us to surpass the state-of-the-art on planar lattices by folding Chignolin (10 amino acids) using the D-Wave 2000Q quantum annealer with a time-to-solution of 0.377$s$.
Furthermore, we have folded the first protein on a cubic lattice using a quantum computer, demonstrating that quantum annealers can reliably fold a fragment of the Trp-Cage protein (8 amino acids) with a time-to-solution of 3.805$s$.

Our work provides the foundation for constructing higher complexity lattice folding models on quantum devices.
These models may support multi-atom representations of amino acids (backbone or sidechain), alternative lattices (body-centered cubic, face-centered cubic, and hexagonal centered cubic), and many-body energy functions.
These advances will ultimately facilitate higher accuracy representation of protein secondary and tertiary structure, and ultimately result in greater utility of lattice models to the problems faced in modern protein design.

In a future study, we will provide empirical scaling comparisons with protein length for the three encodings described in this work and show how the new reverse annealing feature, recently released by D-Wave Systems, can help in decreasing the TTS metric.
Furthermore, we will demonstrate how the ground states of Ising-type Hamiltonians can be obtained using the quantum approximate optimization algorithm (QAOA) by \citet{farhi2014quantum} with hard and soft constraints \citep{qaoa_with_constraints} running on a universal gate-based quantum computer.

\subsection*{Acknowledgments}
We would like to thank Alejandro Perdomo-Ortiz from the NASA Quantum Artificial Intelligence Lab for encouraging us to pursue this line of research.
Special thanks to Yanbo Xue and Aaron Lott at D-Wave Systems Inc.
for continued support and access to the D-Wave 2000Q quantum annealer.
We are grateful for the support of the Rotman School of Business Creative Destruction Lab, especially Daniel Mulet, Hassan Bhatti and Khalid Kurji.
We acknowledge financial support from Data Collective, Spectrum28 and Bloomberg Beta.


\bibliographystyle{apalike}
\bibliography{lattice_proteins_dwave}

\end{document}